\documentclass[showpacs,preprintnumbers,twocolumn]{revtex4}
\usepackage[dvips]{epsfig}

\begin{document}

\preprint{SU-ITP-05-14}

\title{Semi-classical geometry of charged black holes} 
\date{April 8, 2005}

\author{Andrei V. Frolov}
\email{afrolov@stanford.edu}
\affiliation{KIPAC/SITP, Stanford University, Stanford, CA 94305-4060, USA}
\author{Kristj\'an R. Kristj\'ansson}
\email{kristk@hi.is}
\author{L\'arus Thorlacius}
\email{lth@hi.is}
\affiliation{University of Iceland, Science Institute, Dunhaga 3, 
107 Reykjav\'{\i}k, Iceland}

\begin{abstract}
At the classical level, two-dimensional dilaton gravity coupled to an 
abelian gauge field has charged black hole solutions, which have much 
in common with four-dimensional Reissner-Nordstr{\"o}m black holes, 
including multiple asymptotic regions, timelike curvature singularities, 
and Cauchy horizons. The black hole spacetime is, however, significantly 
modified by quantum effects, which can be systematically studied in this 
two-dimensional context. In particular, the back-reaction on the geometry 
due to pair-creation of charged fermions destabilizes the inner horizon 
and replaces it with a spacelike curvature singularity. The semi-classical 
geometry has the same global topology as an electrically neutral black hole.
\end{abstract}

\maketitle

The maximally extended Reissner-Nordstr{\"o}m geometry, describing a 
static electrically charged black hole, has intriguing global structure 
\cite{Graves:1960,Carter:1966}.  There are multiple asymptotic regions 
and Cauchy horizons associated with timelike singularities, which makes 
an initial value formulation problematic. Similar difficulties arise in 
the context of the Kerr spacetime of a rotating black hole. The physical 
relevance of much of the extended structure is questionable, however 
\cite{Penrose:1968,Simpson:1973,McNamara:1978,Chandrasekhar:1982,
Zamorano:1982}. 
At the classical level, a dynamical instability, referred to as mass
inflation, manifests itself when in- and outgoing energy fluxes cross
near the inner horizon, replacing it by an initially null singularity
which turns spacelike deep inside the black hole
\cite{Poisson:1990eh,Ori:1991,Brady:1995ni,Hod:1998gy,Dafermos:2003wr}.
This null singularity is relatively weak, however, with finite integrated 
tidal effects acting on extended timelike observers \cite{Ori:1991}, 
leaving open the possibility of extending the physical spacetime through it. 

It is natural to ask how this classical picture is modified by quantum
effects. These include the pair-creation of charged particles by the 
Schwinger effect \cite{Schwinger:1951nm} in the background electric field 
of the charged black hole. In the Reissner-Nordstr{\"o}m solution the 
electric field diverges as the curvature singularity is approached, 
leading to copious production of electron-positron pairs. At the quantum
level, the black hole charge is screened and the singularity surrounded 
by a charged matter fluid.  This fluid is a source of electric field and 
also modifies the black hole geometry.  The combined electromagnetic and 
gravitational back-reaction can potentially alter the global structure of 
the spacetime \cite{Novikov:1980ni,Herman:1994nv}.  The dynamical 
instability affecting the Cauchy horizon may also be enhanced by the 
production of charged pairs \cite{Anderson:1992pq}. 

The full back-reaction problem is non-trivial and to our knowledge the 
geometry has not been fully elucidated. On the basis of a simple model 
of static electrically charged black holes \cite{Herman:1994nv}, it has
been argued that the effect of pair-production on the interior geometry 
of a black hole can be quite dramatic, in some cases eliminating the 
Cauchy horizon altogether and rendering the singularity spacelike. 
The classical mass inflation instability has been observed in numerical 
work involving the dynamical formation of charged black holes 
\cite{Hod:1998gy}.  In subsequent work \cite{Sorkin:2000pc}, the quantum 
effect of pair-creation was modeled by introducing a non-linear 
dielectric response that prevents the electric field from getting any 
stronger than the critical field for pair-creation. 

In the following we study the internal geometry of electrically charged 
black holes in the simplified context of a 1+1-dimensional model of 
dilaton gravity, coupled to a gauge field and charged Dirac fermions. 
This model has classical charged black hole solutions, whose Penrose 
diagram is identical to that of 3+1-dimensional 
Reissner-Nordstr{\"o}m black holes \cite{McGuigan:1991qp,Frolov:1992xx}.  
Closely related two-dimensional models, with different matter content,
have been shown to exhibit mass inflation at the classical level
\cite{Balbinot:1994ee,Chan:1994tb,Droz:1994aj} and quantum effects due to an 
electrically neutral scalar field were considered in \cite{Balbinot:1994ee}. 
The main new feature of the present work is to include quantum effects 
due to {\it charged\/} matter, which turn out to significantly modify 
the internal structure of a charged black hole. These effects are 
physically important since charged matter fields will be present in any 
model where charged black holes can be formed by gravitational collapse.
In our two-dimensional model, fermion pair-creation is conveniently 
described via bosonization and the resulting effective action allows 
us to study the back-reaction on the geometry in a systematic way.  
Our approach is semi-classical in that it only involves quantum effects 
of the matter field but the dilaton gravity sector remains classical 
throughout. Gravitational quantum effects could also be included along
the lines of \cite{Callan:1992rs,Russo:1992ax, Frolov:1996,Elizalde:1992} 
but we prefer to postpone
that until the effect of charged pair-creation has been mapped out.
We are primarily interested in this 1+1-dimensional 
theory on its own merits as a simplified model of gravity coupled to 
matter but we note that it can be obtained by spherical reduction from 
3+1-dimensional dilaton gravity in the background of an extremal 
magnetically charged black hole 
\cite{Callan:1992rs, Giddings:1992kn,Banks:1992ba}.

The classical action of 1+1-dimensional dilaton gravity coupled to
an abelian gauge field is given by
\begin{equation}
\label{classact}
S_ {dg}  = \int d^2x \sqrt{-g}e^{-2\phi}
\left[R+4(\nabla\phi)^2+4\lambda^2-\frac{1}{4} F^2\right].
\end{equation}
Here $\lambda$ is a dimensionful parameter, inversely related to the charge 
of the 3+1-dimensional extremal dilaton black hole, which sets a mass scale 
for the theory.  In the following we use units where $\lambda=1$.  

The equations of motion of (\ref{classact}) have a two-parameter family 
of static solutions that are analogous to four-dimensional 
Reissner-Nordstr{\"o}m black holes. In the so-called `linear dilaton' gauge, 
they take the form 
\begin{eqnarray}
\label{bhone}
\phi  &=& -x , \qquad F_{01}  = Q e^{-2x} , \\
\label{lindilmetric}
ds^2
&=& -a(x)dt^2+\frac{1}{a(x)}dx^2  , 
\end{eqnarray}
with $a(x)=1-Me^{-2x}+\frac{1}{8}Q^2e^{-4x}$. 
In the asymptotic region
$x\rightarrow\infty$, the metric approaches the two-dimensional Minkowski 
metric and the gravitational coupling strength $e^\phi$ goes to zero. The 
constants $M$ and $Q$ are the mass and charge of the classical geometry. 
Horizons occur at zeroes of the metric function $a(x)$.  We will consider 
non-extremal black holes with $M>\vert Q \vert/\sqrt{2}$, 
for which $a(x)$ has two zeroes at $x=x_\pm$,
\begin{equation} 
e^{2x_\pm}=\frac{1}{2}\left(M\pm\sqrt{M^2-\frac{1}{2}Q^2}\right)
\equiv \psi_\pm .
\label{ypm}
\end{equation} 
The analogy with Reissner-Nordstr{\"o}m black holes can be developed in
detail. The metric (\ref{lindilmetric}) is singular at $x=x_+$ but the 
spacetime curvature is finite there. The singularity signals the breakdown 
of the linear dilaton coordinate system. It is straightforward to find 
coordinates which extend into the region where $\phi>-x_+$. The new
coordinate system, in turn, breaks down at the inner horizon where 
$\phi=-x_-$. The solution can once again be extended but inside the inner 
horizon it eventually runs into a curvature singularity, where the 
gravitational coupling $e^\phi$ diverges. The maximally extended spacetime 
is covered by an infinite number of coordinate patches that occur in a 
repeated pattern. The associated Penrose diagram is identical to that of 
3+1-dimensional Reissner-Nordstr{\"o}m black hole.

Our goal is to determine how the global structure of the black hole
spacetime is modified when Schwinger pair-production is taken into account. 
To this end, we add matter in the form of a 1+1-dimensional Dirac fermion 
to the theory, 
\begin{equation}
\label{matteract}
S_ {m}  = \int d^2x \sqrt{-g}\left[
i\bar\psi\gamma^\mu({\cal D}_\mu+iA_\mu)\psi - m\bar\psi\psi \right].
\end{equation}
With this matter sector, our model can be viewed as a generalization 
to include gravitational effects in the `linear dilaton electrodynamics' 
developed in \cite{Susskind:1992gd,Peet:1992yd,Alford:1992ef}.

The quantum equivalence between the Schwinger model in 1+1 dimensions and 
a bosonic theory with a Sine-Gordon interaction \cite{Coleman:1975pw}
provides an efficient way to include pair-creation effects.
The identification between the fermion field and composite operators 
of a real boson field $Z$ \cite{Coleman:1975pw,Coleman:1976uz} carries 
over to curved spacetime, except for regions of extreme gravity where 
the curvature gets large on the microscopic length scale of the quantum 
theory.  The matter current is given by
\begin{equation}
\label{current}
j^\mu =  \bar\psi\gamma^\mu\psi =
\frac{1}{\sqrt{\pi}}\varepsilon^{\mu\nu}\nabla_\nu Z ,
\end{equation}
where $\varepsilon^{\mu\nu}$ is an antisymmetric tensor, related to
the Levi-Civita tensor density by 
$\varepsilon^{\mu\nu}=\epsilon^{\mu\nu}/\sqrt{-g}$.
The covariant effective action for the boson field is
\begin{equation}
\label{bosonact}
S_ {b}  = \int d^2x \sqrt{-g}\left[-\frac{1}{2} (\nabla Z)^2
- V(Z) -\frac{1}{\sqrt{4\pi}} \varepsilon^{\mu\nu}F_{\mu\nu} Z\right],
\end{equation}
where $V(Z)=c\,m^2 (1-\cos(\sqrt{4\pi}Z))$, with $c$ a numerical constant
whose precise value does not affect our conclusions. 
When the charge-to-mass ratio of the original fermions is large, 
{\it i.e.} when $m \ll 1$ in our units, the semi-classical geometry of a 
charged black hole, including the back-reaction due to pair-creation, is 
reliably described by classical solutions of the combined boson and 
dilaton gravity system, (\ref{classact}) and (\ref{bosonact}).

We work in conformal gauge $ds^2=-e^{2\rho}d\sigma^+d\sigma^-$ and write 
$F^{\mu\nu}=f\varepsilon^{\mu\nu}$ with $f(\sigma^+,\sigma^-)$ a scalar.
The Maxwell equations then reduce to
\begin{equation}
\label{maxwell}
\partial_\pm (e^{-2\phi}f+\frac{1}{\sqrt{\pi}}Z) =0 ,
\end{equation}
and it follows that the gauge field can be eliminated in favor of the 
bosonized matter field
\begin{equation}
\label{zforgauge}
f = -(\frac{1}{\sqrt{\pi}}Z+q)e^{2\phi}.
\end{equation}
The constant of integration $q$ represents a background 
charge located at the strong coupling end of our one-dimensional space. 
In the following we will set $q=0$.  This is natural when we consider 
gravitational collapse of charged matter into an initial vacuum 
configuration. 

Defining new field variables $\psi=e^{-2\phi}$ and $\theta=2(\rho-\phi)$, 
the remaining equations reduce to
\begin{eqnarray}
\label{zeq}
-4\partial_+\partial_-Z
&=& \frac{Ze^\theta}{\pi\psi^2} +
\frac{V'(Z)e^\theta}{\psi} ,\\
\label{psieq}
-4\partial_+\partial_-\psi
&=& (4-\frac{Z^2}{2\pi\psi^2})e^\theta
-\frac{V(Z)e^\theta}{\psi} , \\
\label{thetaeq}
-4\partial_+\partial_-\theta
&=& \frac{Z^2e^\theta}{\pi\psi^3}
+\frac{V(Z)e^\theta}{\psi^2},
\end{eqnarray}
along with two constraints
\begin{equation}
\label{constraints}
\partial_\pm^2\psi-\partial_\pm\theta\partial_\pm\psi 
= -\frac{1}{2}(\partial_\pm Z)^2.
\end{equation}

In the 3+1 dimensional dilaton gravity interpretation, $\psi$ is 
proportional to the area of the transverse two-sphere in the Einstein frame.

\begin{figure}
\begin{tabular}{@{\hspace{-0.5cm}}cc@{\hspace{-0.5cm}}}
  $\psi$\hspace{5em} & $Z$\hspace{5em} \vspace{-1.3em}\\
  \rotatebox{45}{\includegraphics[width=3.1cm]{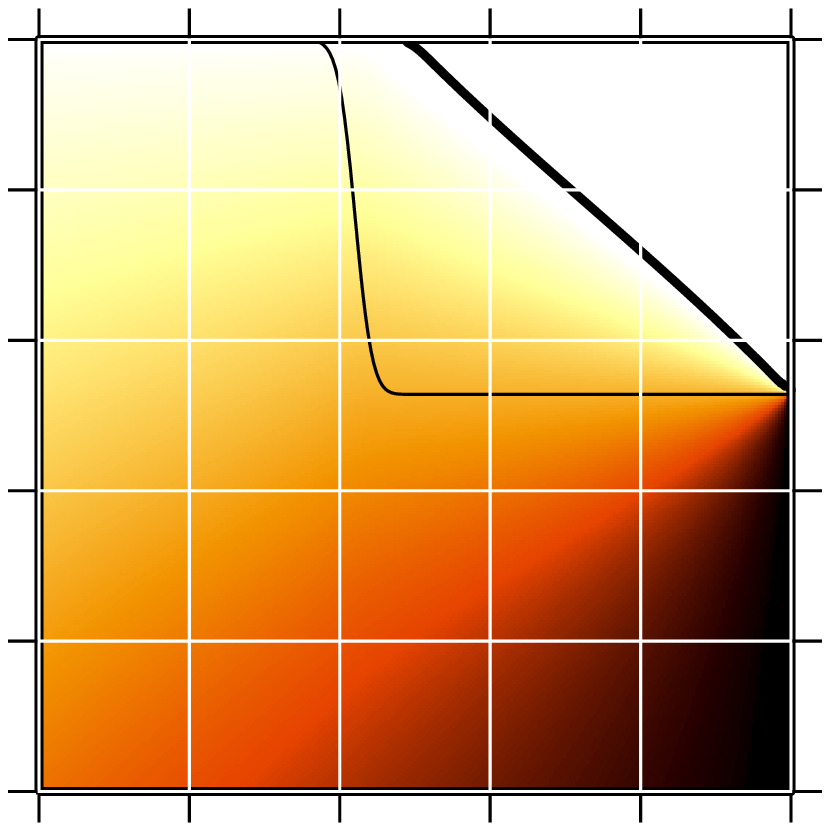}} &
  \rotatebox{45}{\includegraphics[width=3.1cm]{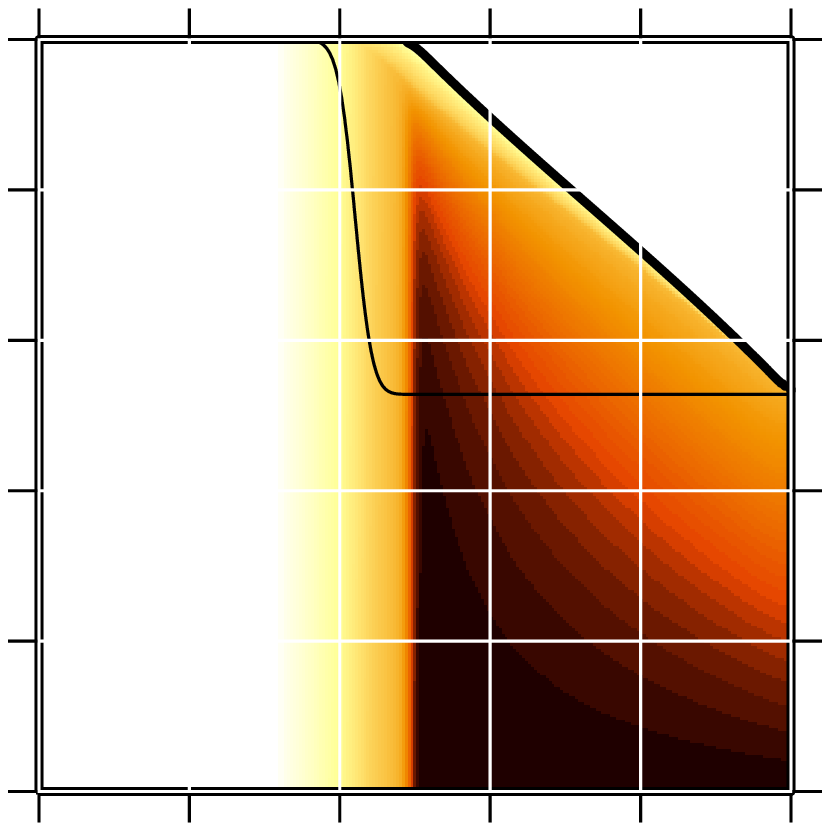}} \vspace{-2em}\\
  \hspace{2.5cm}(a) & \hspace{2.5cm}(b) \\
\end{tabular}
\caption{(a) Density plot of the area function $\psi$ of a black hole 
formed by the gravitational collapse of charged matter. The curves 
indicate the apparent horizon (thin) and the curvature singularity (thick).
(b)~Density plot of the bosonized matter field $Z$. Screening due to 
pair-creation prevents the electric charge from penetrating into the 
strong-coupling region.} 
\label{fig:psi}
\end{figure}

We have solved equations (\ref{zeq})-(\ref{constraints}) numerically on 
a double-null grid using a second-order, staggered leap-frog algorithm. 
A more detailed account will appear in \cite{Frolov:2005} but the main 
points can be summarized as follows.  The numerical evolution requires 
initial data on null slices at early advanced and retarded time 
respectively. For gravitational collapse of charged matter into vacuum 
we describe the incoming matter at early $\sigma^-$ by a kink 
configuration in $Z$ as a function of $\sigma^+$.  The incoming charge 
is given by the height of the kink while the incoming energy is 
determined by its steepness. The slice is chosen to lie at early enough 
$\sigma^-$ so that the incoming matter is initially at weak coupling and 
the black hole has yet to form.  The other initial null slice is taken 
to lie at constant $\sigma^+$ in the vacuum region before the incoming
matter has arrived. This is achieved by choosing the initial kink
configuration such that $\partial_+Z$ has compact support in an interval 
$\sigma^+_0<\sigma^+<\sigma^+_1$.  Then the fields $\psi$, $\theta$, and 
$Z$ take vacuum values on any null slice at constant $\sigma^+<\sigma^+_0$. 

Coordinate transformations that act separately on $\sigma^+$ and $\sigma^-$
preserve the conformal gauge and we use this freedom to fix $\psi$
as some given monotonic function of the coordinates along our initial 
slices. The starting profile of the metric function $\theta$ can then
be obtained by solving the constraint equations (\ref{constraints}) on the
initial slices. Numerical results for $\psi$ and $Z$ obtained from such 
initial data are shown in figure~\ref{fig:psi}. In order to see the global 
geometry we use coordinates that bring $\sigma^+\rightarrow\infty$ to a 
finite distance. The spacetime curvature diverges as $\psi\rightarrow 0$ 
and this occurs on a spacelike curve in figure~\ref{fig:psi}. 
The spacelike singularity approaches the apparent horizon at future null 
infinity, removing all traces of the classical Cauchy horizon inside the 
black hole. We have obtained this behavior for a range of black hole 
masses and charges, and for various (small) values of the fermion mass. 

The semi-classical theory also has static black hole solutions. To study
them we write $\xi=e^\theta$ and define a new spatial coordinate $y$ via 
$dy = \xi d\sigma$. Outside the event horizon the semi-classical equations 
(\ref{zeq})-(\ref{thetaeq}) reduce to 
\begin{eqnarray}
\label{staticzeq}
\xi \ddot{Z} + \dot{\xi}\dot{Z}
&=& \frac{Z}{\pi\psi^2} 
+ \frac{V'(Z)}{\psi} ,\\
\label{staticpsieq}
\xi \ddot{\psi} +\dot{\xi}\dot{\psi}
&=& 4-\frac{Z^2}{2\pi\psi^2}
-\frac{V(Z)}{\psi} , \\
\label{staticxieq}
\ddot{\xi}
&=& \frac{Z^2}{\pi\psi^3}
+\frac{V(Z)}{\psi^2},
\end{eqnarray}
where the dot denotes $\frac{d}{dy}$, and the constraints 
(\ref{constraints}) become 
\begin{equation}
\label{staticconstraint}
\ddot{\psi}+\frac{1}{2}(\dot{Z})^2=0.
\end{equation}
Inside the event horizon $y$ becomes timelike and the 
derivative terms in equations (\ref{staticzeq})-(\ref{staticxieq})
change sign. 

\begin{figure}
\begin{tabular}{c@{\hspace{0.8cm}}c}
\includegraphics[height=4.80cm]{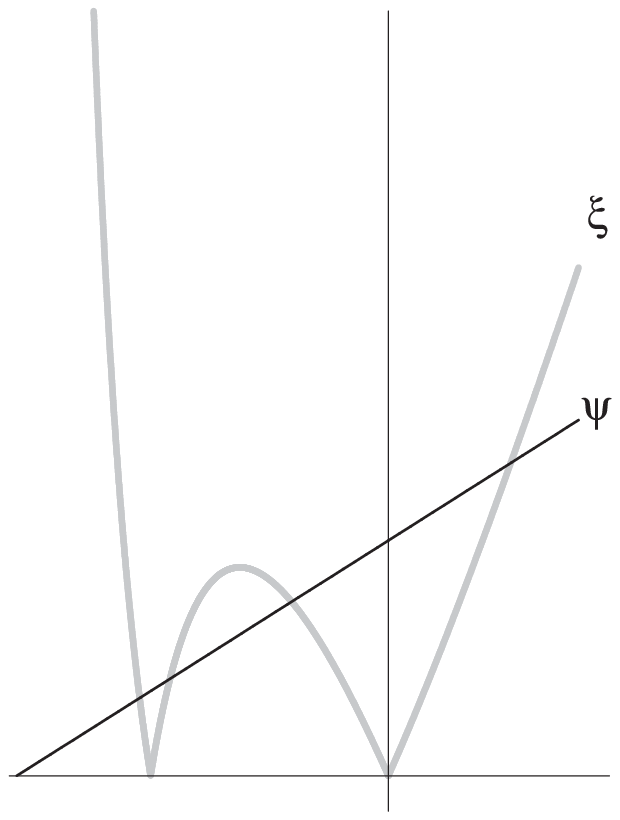} &
\includegraphics[height=4.80cm]{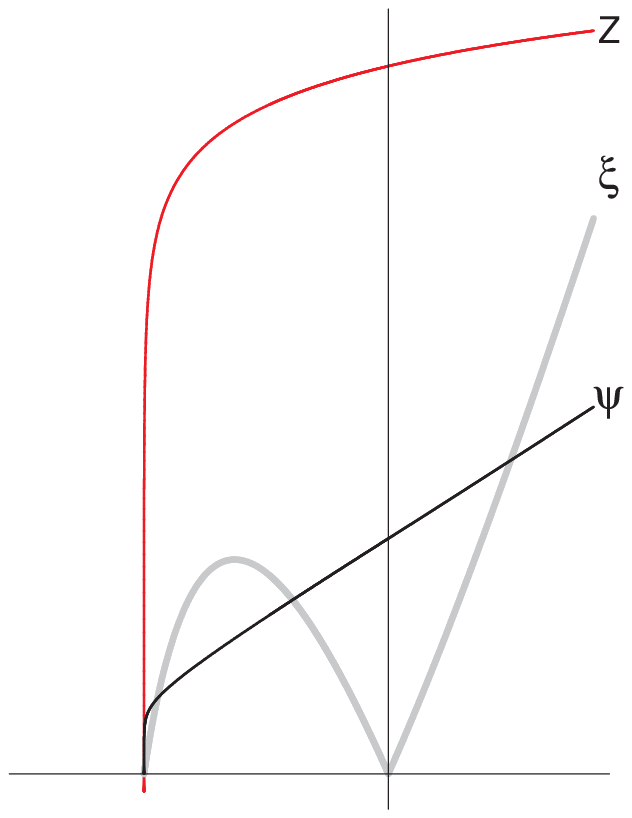} \\
(a) & (b) \\
\end{tabular}
\caption{(a) $\psi$ and $\xi$ plotted as a function of $y$ 
for a classical black hole solution. The two horizons are at the zeroes of 
$\xi$ and the curvature singularity is where $\psi$ goes to zero. 
(b)~Semi-classical black hole. The inner horizon is replaced by a singularity
where $\psi$, $\xi$, and $Z$ all approach zero.}
\label{fig:psixi}
\end{figure}

The corresponding classical system is
\begin{equation}
\label{classicals}
\dot{\xi}\dot{\psi} = 4-\frac{Q^2}{2\psi^2},\qquad
\ddot{\xi} = \frac{Q^2}{\psi^3},\qquad
\ddot{\psi} = 0.
\end{equation}
These equations describe classical black holes of charge $Q$. 
They are obtained from the semi-classical equations by dropping the 
matter field equation of motion (\ref{staticzeq}), and replacing $Z$ in 
equations (\ref{staticpsieq})-(\ref{staticconstraint}) by a constant 
$\sqrt{\pi} Q$, with $Q$ an integer.  In these variables a classical 
black hole solution takes the form
\begin{eqnarray}
\psi(y)
&=& \psi_+ + \alpha y,\nonumber \\ 
\label{classext}
\xi(y)
&=& \frac{4}{\alpha^2}
\left\vert \alpha y+\frac{\psi_+\psi_-}{\psi_++\alpha y}-\psi_-\right\vert ,
\end{eqnarray}
where $Q^2=8\psi_+\psi_-$ has been used. The free parameter $\alpha$ sets 
the scale of the spatial coordinate $y$, with $\alpha=2$ matching the
coordinate scale in (\ref{bhone})-(\ref{lindilmetric}). The solution 
(\ref{classext}) is shown in figure~\ref{fig:psixi}(a). It describes all 
three regions: outside, inside, and between the two horizons.  
The area function $\psi$ extends smoothly through both horizons. 
It decreases linearly as we go deeper into the black hole and reaches 
zero at the curvature singularity. Meanwhile $\xi$, which contains the 
conformal factor of the metric, goes to zero at both horizons and diverges 
at the singularity. The absolute value sign in (\ref{classext}) reflects 
the signature change of the metric between horizons. 

Returning to the semi-classical equations 
(\ref{staticzeq})-(\ref{staticconstraint}) we find significant departure
from the above classical behavior. Explicit analytic solutions are not 
available but the equations can be integrated numerically. A black hole 
solution with a smooth event horizon is obtained by starting the 
integration at the horizon $y=0$ with $\xi(0)=0$, and tuning the initial 
values $\dot{\psi}(0)$ and $\dot{Z}(0)$ so that the solution is regular 
there \cite{Birnir:1992by}.  Different black holes are then parametrized 
by $\psi(0)$ and $Z(0)$. The choice of $\dot{\xi}(0)$ sets the scale of 
$y$ but does not affect the geometry.

Numerical results for massless matter are shown in 
figure~\ref{fig:psixi}(b). A small fermion mass does not change the
qualitative behavior.
The scalar fields $\psi$ and $Z$ extend smoothly through the 
horizon at $y=0$, while $\xi$ goes to zero there and $\dot{\xi}$ changes 
sign, as in the classical solution (\ref{classext}).  The new features
emerge when we follow the semi-classical solution inwards from $y=0$. 
The amplitude of the matter field decreases as we go into the black hole. 
This is due to pair-creation and it causes the area function $\psi$ to 
decrease more rapidly than the linear behavior of the classical solution.
In fact the constraint equation (\ref{staticconstraint}) implies that
$\psi$ is a concave function and any variation in the matter field will
focus it towards zero. By equation (\ref{staticxieq}) the metric function 
$\xi$ is also concave and approaches zero.  At a smooth inner horizon the 
fields $\psi$ and $Z$, along with their derivatives, would remain finite 
as $\xi$ goes to zero. If, on the other hand, $\psi$ goes to zero the
gravitational sector becomes infinitely strongly coupled and we expect
a curvature singularity.
In our numerical evaluation all the fields $\xi$, $\psi$ and $Z$ are 
simultaneously driven to zero while $\dot{\xi}$, $\dot{\psi}$, and 
$\dot{Z}$ become large. The Ricci scalar increases rapidly as the 
singular point is approached. This strongly suggests that the classical 
inner horizon is replaced by a curvature singularity at the 
semi-classical level. Unfortunately the numerical solutions are not 
detailed enough to describe the final approach to the singularity, but 
the numerical evidence indicates that it is spacelike, which is also
what we found in the gravitational collapse solutions.

This work was supported in part by the Institute for Theoretical Physics
at Stanford University and by grants from the Icelandic Science and 
Technology Policy Council and the University of Iceland Research Fund.

\end{document}